\documentclass[aps,prl,preprint,groupedaddress,showpacs]{revtex4}

\usepackage{graphicx}
\usepackage{dcolumn}
\usepackage{bm}

\begin{document}


\title{Optical-fiber source of polarization-entangled photon pairs
in the 1550\,nm telecom band}



\author{Xiaoying Li, Paul L. Voss, Jay E. Sharping, and Prem Kumar}
 \email[]{xyli@ece.northwestern.edu}
\affiliation{Center for Photonic Communication and Computing, ECE
Department \\ Northwestern University, Evanston, IL 60208-3118}


\date{August 9, 2004}

\begin{abstract}
We present a fiber based source of polarization-entangled photon
pairs that is well suited for quantum communication applications
in the 1550\,nm band of standard fiber-optic telecommunications.
Polarization entanglement is created by pumping a nonlinear-fiber
Sagnac interferometer with two time-delayed orthogonally-polarized
pump pulses and subsequently removing the time distinguishability
by passing the parametrically scattered signal and idler photon
pairs through a piece of birefringent fiber. Coincidence detection
of the signal and idler photons yields biphoton interference with
visibility greater than 90\%, while no interference is observed in
direct detection of either the signal or the idler photons. All
four Bell states can be prepared with our setup and we demonstrate
violations of the CHSH form of Bell's inequality by up to 10
standard deviations of measurement uncertainty.
\end{abstract}


\maketitle


Quantum entanglement refers to the nonclassical interdependency of
physically separable quantum subsystems. In addition to being at
the heart of the most fundamental tests of quantum
mechanics~\cite{Einstein35,Bell64,Clauser69,Greenberger90}, it is
an essential resource that must be freely available for
implementing many of the novel functions of quantum information
processing~\cite{Bennett92,Bennett93}. In photonic systems, the
ongoing developments in lasers, optical-fiber technology,
single-photon detectors, and nonlinear optical materials have led
to enormous experimental progress in both the
fundamental~\cite{Aspect82,Ou88,Ou90,Shih88,Pan00} and applied
domains~\cite{Mattle96,Bouwmeester97,Gisin02}. A popular approach to
generating entangled pairs of photons is based on the nonlinear
process of parametric down conversion in $\chi^{(2)}$
crystals~\cite{Kwiat95,Rarity90,Brendel92}. Though much progress
has been made using this approach, formidable engineering problems
remain in coupling the entangled photons into standard optical
fibers for transmission, storage, and manipulation over long
distances.

The coupling problem can be obviated if the entangled photons can
be generated in the fiber itself, and desirably, in the fiber's
low-loss propagation window near 1.5\,$\mu$m, since that would
minimize losses during transmission as well. Apart from the
inherent compatibility with the transmission medium, a fiber based
source of entangled photons would have other advantages over its
crystal
counterparts~\cite{Kwiat95,Rarity90,Brendel92,Sanaka01,Kim99,Kurtsiefer00}.
Particularly, the spatial mode of the photon-pair would be the
guided transverse mode of the fiber, which is a very pure
Gaussian-like single spatial mode in modern fibers. A well-defined
mode is highly desirable for realizing complex networks involving
several entangling operations. In this Letter, we describe the
first, to the best of our knowledge, optical fiber source of
polarization-entangled photon pairs in the 1550\,nm telecom band.
A variety of bi-photon interference experiments are presented that
show the nature of the entanglement generated with this source.
All four Bell states can be prepared with our setup and the CHSH
form of Bell's inequality is violated by up to 10 standard
deviations of measurement uncertainty.

\begin{figure}
\includegraphics[width=3in]{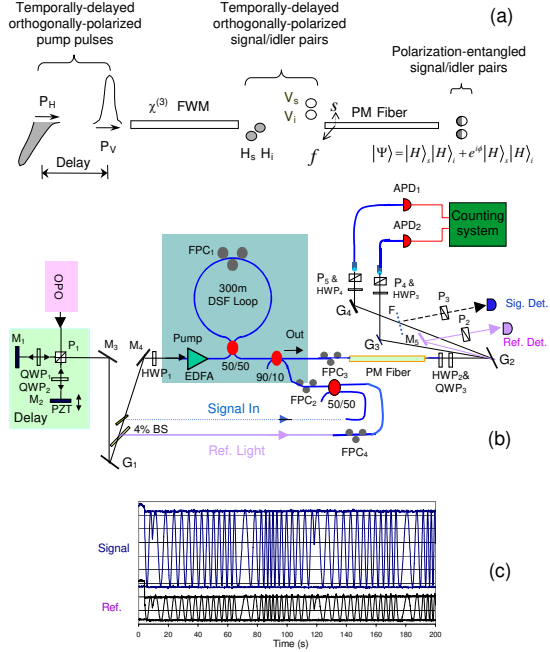}
\caption{\label{setup-v2}(a) Conceptual representation of the
multiplexing scheme used in our experiment to create
polarization-entangled pairs of photons. (b) Schematic of the
experimental setup. P$_1$--P$_5$, polarization beam splitters;
G$_1$--G$_4$, diffraction gratings; M$_1$--M$_5$, mirrors;
FPC$_1$--FPC$_4$, fiber polarization controllers; QWP,
quarter-wave plate; HWP, half-wave plate; F, flipper mirror. (c)
Sinusoidal variations (or constancy at the peaks and troughs) of
the photocurrents obtained from the signal (top traces) and the
reference detectors (bottom traces) upon linearly sweeping the
voltage (or maintaining a fixed voltage) on the PZT. The clarity
of the traces demonstrates minutes-long stability of the
polarization interferometer formed between P$_1$ and P$_3$ (P$_2$)
for signal (reference) light.}
\end{figure}

Recently, our group has demonstrated that parametric fluorescence
accompanying nondegenerate four-wave mixing (FWM) in standard
optical fibers is an excellent source of quantum-correlated photon
pairs~\cite{Fiorentino02a}. The quantum correlation arises from
four-photon scattering (FPS) events, wherein two pump photons at
frequency $\omega_{p}$ scatter through the Kerr nonlinearity of
the fiber to simultaneously create a signal photon and an idler
photon at frequencies $\omega_{s}$ and $\omega_{i}$, respectively,
such that $\omega_s + \omega_i = 2\omega_p$. For a linearly
polarized pump with wavelength close to the zero-dispersion
wavelength of the fiber, the FWM process is phase-matched and the
accompanying parametric fluorescence is predominantly co-polarized
with the pump. Since the response time of the Kerr nonlinearity is
almost instantaneous, two such parametric scattering processes can
be time and polarization multiplexed to create the desired
polarization entanglement. For example [see
Fig.~\ref{setup-v2}(a)], when the fiber is pumped with two
orthogonally polarized, relatively delayed pulses, the
signal-idler photon pairs scattered from each pulse are
co-polarized with that pump pulse and relatively delayed by the
same amount.  The distinguishing
time delay between the orthogonally-polarized photon pairs,
however, can be removed by passing the pairs
through a piece of birefringent fiber of appropriate length,
wherein the photon-pair travelling along the fast axis of the
fiber catches up with the other pair travelling along the slow
axis. When the emerging signal and idler photons are separated
based on their wavelength, each stream of photons is completely
unpolarized because any polarizer/detector combination is unable
to determine which pump pulse a detected photon originated from.
When the signal and idler photons are detected in coincidence, it
is still impossible to determine which pump pulse created the
detected pair. This indistinguishability gives rise to
polarization entanglement in our experiment.

A schematic of the experimental setup is shown in
Fig.~\ref{setup-v2}(b). Signal and idler photon pairs at
wavelengths of 1547.1\,nm and 1525.1\,nm, respectively, are
produced in a nonlinear fiber Sagnac interferometer
(NFSI)~\cite{Fiorentino02a,Sharping01}. The NFSI consists of a
fused-silica 50/50 fiber coupler spliced to 300\,m of
dispersion-shifted fiber (DSF) that has a zero-dispersion
wavelength at $\lambda_0=1535\pm 2$\,nm.  Because the Kerr
nonlinearity is weak, for this length of fiber only about 0.1
photon-pair is produced with a typical 5\,ps-duration pump pulse
containing $\sim$$10^7$ photons. Thus, to reliably detect the
correlated photon pairs, a pump-to-signal rejection ratio in
excess of 100\,dB is required. We achieve this by first exploiting
the mirror-like property of the Sagnac loop, which provides a pump
rejection of $>30$\,dB, and then sending the transmitted
fluorescence photons along with the leaked pump photons through a
free-space double-grating spectral filter (DGSF) that provides a
pump rejection ratio in excess of 75\,dB~\cite{Fiorentino02a}. The
filter consists of three identical diffraction gratings
(holographic, 600 grooves/mm) G$_2$, G$_3$, and G$_4$, whose
diffraction efficiencies for the horizontally and the vertically
polarized light are 90\% and 86\%, respectively. The
doubly-diffracted signal and idler photons are then re-coupled
into fibers, whose numerical apertures along with the geometrical
settings of the gratings determine the pass bands for the signal
and idler channels. The full-width at half-maximum (FWHM)
bandwidth for both the channels is 0.6\,nm.

During the experiment, for alignment and phase control purposes,
input-signal and reference pulses are also needed that are
temporally synchronized with the pump pulses. The main purpose of
the signal pulses is to ensure that the time distinguishability
between the orthogonally-polarized photon-pairs is effectively
removed. By spectrally carving~\cite{Sharping01} the
$\sim$150\,fs-pulse train from an optical parametric oscillator
(Coherent Inc., model Mira-OPO), we obtain trains of 5-ps pump
pulses with central wavelength at 1536\,nm, 2.8-ps signal pulses
with central wavelength at 1547\,nm, and 4-ps reference pulses
with central wavelength at 1539\,nm. The pump pulses are then
amplified by an erbium-doped fiber amplifier (EDFA) to achieve the
required average pump power. Light at the signal and idler
wavelengths from the OPO that leaks through the spectral-carving
optics and the amplified spontaneous emission from the EDFA are
suppressed by passing the pump pulses through a 1\,nm-bandwidth
tunable optical filter (Newport, TBF-1550-1.0).

A 30-ps relative delay between the two orthogonally-polarized pump
pulses is introduced by adding separate free-space propagation
paths for the two pulses with use of a polarization beam splitter
(PBS) P$_1$, quarter wave-plates (QWP) QWP$_1$ and QWP$_2$, and
mirrors M$_1$ and M$_2$. Mirror M$_2$ is mounted on a
piezoelectric-transducer-(PZT)-driven translation stage, which
allows precise adjustment of the relative delay and phase
difference between the orthogonally-polarized pump-pulse pairs.
After the NFSI, the delay is compensated by propagating the
scattered photon pairs along the fast and slow polarization axes
of 20\,m-long polarization-maintaining (PM) fiber. A careful
alignment procedure is implemented to properly orient the axes of
the PM fiber, taking into consideration the change of polarization
state incurred by an input signal-pulse pair upon
maximally-amplified reflection from the NFSI~\cite{Mortimore88}.
Alignment is performed prior to the actual experiment by injecting
weak path-matched signal-pulse pairs, having identical temporal
and polarization structure as the pump pulses, into the NFSI
through the 50/50 and 90/10 couplers. First the signal
amplification is maximized by adjusting FPC$_2$, while monitoring
the signal gain on a detector (ETX500) placed after P$_3$. Then
the fringe visibility of the polarization interferometer formed
between P$_1$ and P$_3$ is maximized by adjusting FPC$_3$, HW$_2$,
and QWP$_3$ while observing the fringes in real time upon periodic
scanning of M$_2$. Once the alignment is completed, the injected
signal is blocked and further measurements are made only on the
parametric fluorescence.

After compensation of the time delay, the following
polarization-entangled state is generated at the output of the PM
fiber: $|\Psi\rangle=|H\rangle_s |H\rangle_i + e^{i \phi}
|V\rangle_s |V\rangle_i$, where $\phi$ is the relative phase
difference between the two-photon amplitudes $|H\rangle_s
|H\rangle_i$ and $|V\rangle_s |V\rangle_i$. In our experiment
$\phi = 2\phi_p$ with $\phi_p$ being the relative phase difference
between the two delayed, orthogonally-polarized pump pulses. This
source can produce all four polarization-entangled Bell states.
When $\phi_p=0, \frac{\pi}{2}$, the states $|\Psi^\pm \rangle =
|H\rangle_s |H\rangle_i \pm |V\rangle_s |V\rangle_i$ are created.
The other two Bell states $|\Phi^\pm\rangle = |H\rangle_s
|V\rangle_i \pm |V\rangle_s |H\rangle_i$ can be prepared by
inserting a properly oriented HWP in the idler channel.
Non-maximally entangled pure states with an arbitrary degree of
polarization entanglement can also be created with our setup by
choosing the two pump pulses to have unequal powers.

In order to actively monitor and control the relative phase $\phi$
during the course of data taking, weak reference-pulse pairs of
about 50\,$\mu$W average power are injected into the NFSI through
the 50/50 and 90/10 couplers. The reference-pulse pairs have
identical temporal and polarization structure as the pump pulses,
except the temporal location of the reference-pulse pairs is
mismatched with respect to the pump-pulse pairs and their
wavelength is slightly detuned, so that they neither interact with
the pump pulses nor are seen by the single-photon detectors used
in the signal and idler channels. During the course of
measurements on the polarization-entangled states, the relative
phase between the reference-pulse pairs, $\phi_{\rm ref}$, is
monitored by measuring the photocurrent from a low-bandwidth
reference detector placed after P$_2$ to make observations on one
output port of the polarization interferometer [see Fig.~1(b)].
The voltage created by this photocurrent is compared to a
reference voltage and the difference is used to stabilize
$\phi_{\rm ref}$ by feeding back on the PZT through an electronic
circuit. The excellent overall stability of the system is shown by
the near-perfect classical interference fringes displayed in the
inset in Fig.~\ref{setup-v2}(b), which were simultaneously
obtained with injected signal light and with reference light while
scanning $\phi_{\rm ref}$ by ramping the voltage on the PZT. The
relative phase between the reference-pulse pairs, $\phi_{\rm
ref}$, is related to the relative phase between the pump-pulse
pairs via $\phi_p = \phi_{\rm ref} + \delta$, where $\delta$
results from dispersion in the DSF owing to slightly different
wavelengths of the pulse pairs.

\begin{figure}
\includegraphics[width=2.5in]{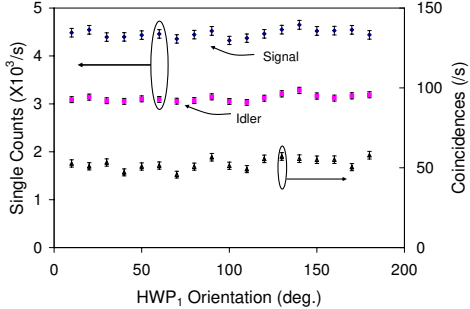}
\caption{\label{fig02-v2} Observed polarization (in)dependence of
parametric fluorescence in the DSF.}
\end{figure}

The photon-counting modules used for detecting the signal and
idler photons consist of InGaAs/InP avalanche photodiodes (APDs,
Epitaxx, EPM 239BA) operated in a gated-Geiger
mode~\cite{Fiorentino02a}. Gate pulses of 1\,-ns duration are
applied at 588\,kHz rate, 1/128 of the pump-pulse rate, to avoid
after-pulsing in the detectors. The measured quantum efficiencies
for the two detectors are 25\% and 20\%, respectively. The overall
detection efficiencies for the signal and idler photons are about
9\% and 7\%, respectively, when the transmittance of the Sagnac
loop (82\%), 90/10 coupler, DGSF (57\%), and other optical
components (90\%) are included. Given a parametric scattering
probability of $\simeq$0.1 pairs/pulse in the DSF, corresponding
to 0.39\,mW of average pump power in each direction around the
Sagnac loop, and the gate rate of 588\,kHz, we typically observe
$\simeq$4000 counts/s in the signal and idler channels when
detecting the parametric fluorescence.

The polarization-entanglement generation scheme described here
uses the fact that the FPS efficiency does not depend on the
pump-polarization direction. Moreover, the signal and idler
fluorescence photons are predominantly co-polarized with the pump.
Although the Kerr susceptibility tensor for isotropic fused-silica
glass gives a cross-polarized scattering probability that is only
1/9th as strong, the actual scattering probability for
cross-polarized fluorescence is even weaker owing to differing
phase-matching condition for cross-polarized photons. We verify
the polarization independence of the FPS efficiency in the fiber
by monitoring the parametric fluorescence while varying the
polarization direction of the injected pump pulses with use of a
half-wave plate (HWP$_1$). The individual counts for the signal
and idler photons, and their coincidence counts, versus the
HWP$_1$ angle are shown in Fig.~\ref{fig02-v2}. The slight
variation observed in the count rates is due to
polarization-dependent transmission of the DGSF. Note that for the
measurements shown in Fig.~\ref{fig02-v2} the input pump delay,
the PM-fiber delay compensation, and the detection analyzers were
removed.

Polarization correlations are measured by inserting adjustable
analyzers in the paths of signal and idler photons, each
consisting of a PBS (P$_4$, P$_5$) preceded by an adjustable HWP
(HWP$_3$, HWP$_4$). For the state $|\Phi\rangle = |H\rangle_s
|V\rangle_i + e^{i \phi}|V\rangle_s |H\rangle_i$, when the
relative phase between the down-converted pairs is $\phi$ and the
polarization analyzers in the signal and idler channels are set to
$\theta_1$ and $\theta_2$, respectively, the single-count
probability for the signal and idler photons is $R_i=\alpha_i/2$
($i=1,2$) and the coincidence-count probability R$_{12}$ is given
by
\begin{eqnarray}
R_{12}=2^{-1}\alpha_1 \alpha_2 [\sin^2 \theta_1 \cos^2 \theta_2 +
\cos^2 \theta_1 \sin^2 \theta_2  \nonumber \\ + 2 \cos\phi \sin
\theta_1 \cos \theta_1 \sin \theta_2 \cos \theta_2], \label{eq1}
\end{eqnarray}
where $\alpha_i$ is the total detection efficiency in the two
channels.

\begin{figure}
\includegraphics[width=2.5in]{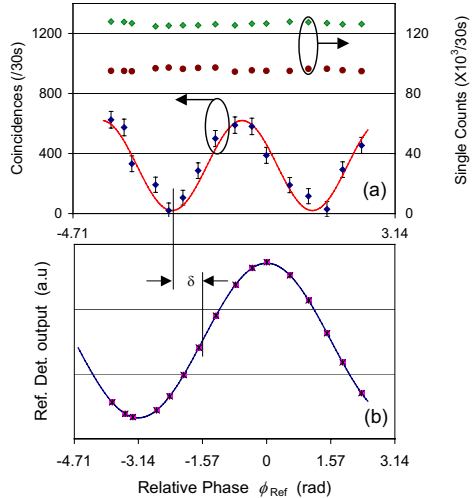}
\caption{\label{fig03-v2}(a) Coincidence counts and single counts
detected over 30\,s when the relative phase $\phi_{\rm ref}$ is
varied. The solid curve a fit to Eq.~\ref{eq1}. (b) Output from
the reference detector versus $\phi_{\rm ref}$ showing the
ordinary one-photon interference with twice the fringe spacing as
in (a). }
\end{figure}

We performed three sets of experiments to evaluate the degree of
polarization entanglement of our source. The first measurement
consisted of setting both analyzers at 45$^\circ$ and slowly
scanning $\phi_{\rm ref}$ by applying a voltage ramp on the PZT.
As shown in Fig.~\ref{fig03-v2}(a), the coincidence counts reveal
sinusoidal variation with a fringe visibility of 93\% (dark counts
and accidental-coincidence counts have been subtracted), while the
single counts remain unchanged. The output from the reference
detector is also recorded simultaneously, which is shown in
Fig.~\ref{fig03-v2}(b). The relative shift of the sinusoidal
variation of two-photon interference in Fig.~\ref{fig03-v2}(a)
from that of reference-light interference in
Fig.~\ref{fig03-v2}(b) is a direct measure of the phase shift
$\delta$, which is used below to properly set $\phi$ for
measurements of the violation of Bell's inequality.

\begin{figure}
\includegraphics[width=2.5in]{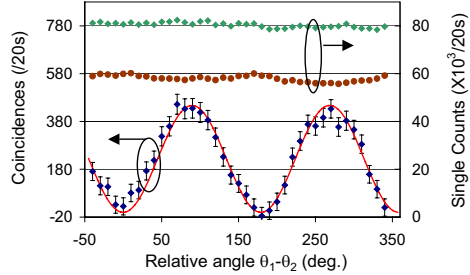}
\caption{\label{fig04-v2} Measurement of polarization
entanglement: Coincidence counts and single counts detected over
20\,s as the analyzer angle in the idler channel is varied while
keeping the signal-channel analyzer fixed at 45$^\circ$ relative
to vertical.}
\end{figure}

In the second set of measurements on polarization entanglement, we
locked the generated state to $|\Phi^-\rangle=|H\rangle_s
|V\rangle_i-|V\rangle_s |H\rangle_i$ by applying an appropriate
feedback on the PZT, fixed the angle of the polarization analyzer
in the signal channel to 45$^\circ$, and varied the analyzer angle
in the idler channel by rotating HWP$_4$. The result is shown in
Fig.~\ref{fig04-v2}. As expected, the coincidence-count rate
displays sinusoidal interference fringes with a visibility of
92\%, whereas the variation in the single-count rate is only 4\%
(once again, dark counts and accidental coincidences have been
subtracted).

In the third set of experiments, we characterized the quality of
polarization entanglement produced with our source through
measurements of Bell's inequality violation. By recording
coincidence counts for 16 different combinations of analyzer
settings with $\theta_1=0^\circ, 90^\circ, -45^\circ, 45^\circ$
and $\theta_2=-22.5^\circ, 67.5^\circ, 22.5^\circ, 112.5^\circ$,
we measured the quantity $S$ in the CHSH form of Bell's
inequality~\cite{Clauser69}, which satisfies $|S|\leq 2$ for any
local realistic description of our experiment. The results, which
are presented in Table~\ref{bellviolation}, show that a) the CHSH
inequality is violated, i.e., $|S|>2$, for all four Bell states
produced with our setup and b) the violation occurs by up to 10
standard deviations of measurement uncertainty.

\begin{table}
\caption{\label{bellviolation} Measured values of $S$ for the four
Bell states.}
\begin{ruledtabular}
\begin{tabular}{ccc}
 Bell state & $S$ & Violation \\
 & & (standard deviations) \\
\hline
 $|H\rangle_s |H\rangle_i + |V\rangle_s |V\rangle_i$
 & $2.75 \pm 0.077$ & 10\,$\sigma$ \\
 $|H\rangle_s |H\rangle_i - |V\rangle_s |V\rangle_i$
 & $2.55 \pm 0.070$ & 8\,$\sigma$ \\
 $|H\rangle_s |V\rangle_i + |V\rangle_s |H\rangle_i$
 & $2.48 \pm 0.078$ & 6\,$\sigma$ \\
 $|H\rangle_s |V\rangle_i - |V\rangle_s |H\rangle_i$
 & $2.64 \pm 0.076$ & 8\,$\sigma$ \\
\end{tabular}
\end{ruledtabular}
\end{table}

In order to ascertain the degree of entanglement produced by the
true FPS events in our setup, the accidental coincidences
resulting from the uncorrelated background photons and the dark
counts in the detectors were measured for each set of data
acquired in the three polarization-entanglement experiments
described above. The rate of accidental coincidences was as large
as the rate of ``true" coincidences plotted in
Figs.~\ref{fig02-v2}--\ref{fig04-v2} by subtracting the accidental
coincidences and the raw visibility of two-photon interference was
only $\simeq 30$\%. We believe the majority of background photons
in our setup arise from spontaneous Raman
scattering~\cite{Voss04-OL,Voss04-JOB}. Our recent measurements
with a modified DGSF have shown that the contribution of
accidental coincidences can be made $<10\%$ of the total measured
coincidences~\cite{Li04-OpEx}. With these improvements, a raw
two-photon-interference visibility of $>85\%$ would be obtained,
i.e., without any post-measurement corrections.

In conclusion, we have developed and characterized a fiber-based
source of polarization-entangled photon pairs. The count rates in
the experiment at present are limited by the repetition rate of
the APDs, which can be increased by at least an order of magnitude
by straightforward refinements of the detection electronics. The
photon-pair production rate, on the other hand, is limited by the
75\,MHz repetition rate of the pump laser. The production rate can
be dramatically increased by using state-of-the-art pulsed lasers
that have been developed for fiber-optic communications. These
lasers operate at 10--40\,GHz repetition rates and can have the
requisite peak-pulse powers with use of medium-power EDFAs.
Bulk-optic implementations of the pump delay apparatus and the
detection filters were used in these proof-of-principle
experiments for purposes of tunability and control. All-fiber
versions of these subsystems can be readily realized with use of
PM fibers, wavelength-division-multiplexing filters, and fiber
polarizers. Finally, we have understood the origin of the large
number of accidental coincidences in the experiment and subsequent
system improvements are expected to significantly improve the
degree of entanglement produced with our system. Therefore, we
believe that such fiber-based entangled-photon pairs will prove to
be an efficient source for developing quantum communication
technologies.

\begin{acknowledgments}
This work was supported in part by the DoD Multidisciplinary
University Research Initiative (MURI) Program administered by the
Army Research Office under Grant DAAD19-00-1-0177.
\end{acknowledgments}

\newcommand{\noopsort}[1]{} \newcommand{\printfirst}[2]{#1}
  \newcommand{\singleletter}[1]{#1} \newcommand{\switchargs}[2]{#2#1}

\end{document}